\documentclass[reprint,showpacs,amsmath,amssymb,aps,prb]{revtex4-1}

\usepackage{bm}
\usepackage{dcolumn}
\usepackage{graphicx}
\usepackage{subfig}
\usepackage{multirow}

\begin{document}

\title{Electronic structure and phase stability of oxide semiconductors: performance of  dielectric-dependent hybrid functional DFT, benchmarked against $\mathit{GW}$ bandstructure calculations and experiments}

\author{Matteo Gerosa}
\affiliation{Department of Energy, Politecnico di Milano, via Ponzio 34/3, 20133 Milano, Italy}
\author{Carlo Enrico Bottani}
\email[Corresponding author: ]{carlo.bottani@polimi.it}
\affiliation{Department of Energy, Politecnico di Milano, via Ponzio 34/3, 20133 Milano, Italy}
\affiliation{Center for Nano Science and Technology @Polimi, Istituto Italiano di Tecnologia, via Pascoli 70/3, 20133 Milano, Italy}
\author{Lucia Caramella}
\author{Giovanni Onida}
\affiliation{Dipartimento di Fisica, Universit\`a degli Studi di Milano, Milano, Italy}
\affiliation{European Theoretical Spectroscopy Facility (ETSF)}
\author{Cristiana Di Valentin}
\author{Gianfranco Pacchioni}
\affiliation{Dipartimento di Scienza dei Materiali, Universit\`a di Milano-Bicocca, via R. Cozzi 55, 20125 Milan, Italy}

\date{\today}%

\begin{abstract}
 We investigate band gaps, equilibrium structures and phase stabilities of several bulk polymorphs of wide gap oxide semiconductors ZnO, TiO$_2$, ZrO$_2$ and WO$_3$. We are particularly concerned with assessing the performance of hybrid functionals built with the fraction of Hartree-Fock exact exchange obtained from the computed electronic dielectric constant of the material. We provide comparison with more standard density-functional theory and $\mathit{GW}$ methods. We finally analyze the chemical reduction of TiO$_2$ into Ti$_2$O$_3$, involving a change in oxide stoichiometry. We show that the dielectric-dependent hybrid functional is generally good at reproducing both ground-state (lattice constants, phase stability sequences and reaction energies) and excited-state (photoemission gaps) properties within a single, fully \emph{ab initio} framework.
\end{abstract}

\maketitle

\section{Introduction}
\label{sec:introduction}

Density-functional theory (DFT) has become the method of choice for calculating, understanding, and possibly predicting properties of real materials from first-principles. The approach developed by Kohn and Sham\cite{kohn1965} (KS)  provides a  computational framework that allows to calculate ground-state related properties of realistic systems at affordable computational cost. For example, lattice constants of materials are often obtained in excellent agreement with experiment. In contrast, excited states are usually poorly described, being beyond the scope of standard DFT-KS. A prototypical exemplification of this behavior is provided by the well-known failure of local and semilocal energy functionals in estimating the electronic band gap of semiconductors and insulators. In fact, the electronic gap obtained in the KS scheme (the KS gap) differs from true quasiparticle gap (measured e.g. in photoemission and inverse photoemission experiments) by the derivative discontinuity of the exchange-correlation (xc) 
energy functional at hand.\cite{perdew1983,* sham1983}
A rigorous description of electronic excitations requires working in the framework of many-body perturbation theory.\cite{onida2002} Within the latter, the $\mathit{GW}$ approximation\cite{hedin1965} is to date the state-of-the-art method for computing quasiparticle band structures that can be meaningfully compared with experiments. However, $\mathit{GW}$ calculations are much more computationally demanding than DFT-based ones, indeed limiting the size of the systems that can be investigated at reasonable cost. Although methodological advances have recently made it possible to treat remarkably complex surface/interface systems at the $\mathit{GW}$ level,\cite{umari2013,verdi2014} DFT is still widely used for investigating electronic structure features in bulk solids and surfaces, for which in most cases it is able to provide at least a qualitative picture. This can be justified on the basis of the formal similarity between the KS equation and the equation for the poles of the 
Green's function of many-body perturbation theory, once the xc potential is identified with an approximation of the self-energy operator $\Sigma$. The development of hybrid functionals,\cite{becke1993a,perdew1996-pbe0,stephens1994,becke1993b} first proposed within the chemistry community with the aim of increasing accuracy in molecular calculations and now routinely applied also to extended, condensed systems,\cite{muscat2001, cora2004,marsman2008} has allowed to gain even quantitative accuracy in the computation of electronic band structures. 

Hybrid functionals are constructed by admixing a portion of Hartree-Fock exact exchange to a semilocal xc potential within the generalized-gradient approximation (GGA). Built upon a generalized KS scheme, the hybrid functional approach provides a practical way to cure some of the known drawbacks of local and semilocal DFT xc functionals,\cite{kummel2008} and constitutes an alternative, cheaper electronic structure method with respect to $\mathit{GW}$. This is especially true if localized basis sets are used for evaluating exact exchange, which allows to reduce the computational effort down to that of standard calculations within the local-density approximation (LDA) or the GGA. Instead, using plane waves makes calculations about two order of magnitude more expensive, again limiting the size of the systems able to be addressed.
Standard recipes for hybrid functionals prescribe a \emph{fixed} portion $\alpha$ of exact exchange, independently of the system being investigated. This is the case of the full-range Perdew-Burke-Ernzerhof (PBE0)\cite{perdew1996-pbe0} and screened-exchange Heyd-Scuseria-Ernzerhof (HSE06)\cite{heyd2003,* heyd2006} hybrid functionals, in which $\alpha$ is set to 1/4. The latter has been identified as the optimal value for obtaining accurate atomization energies in molecular calculations.\cite{perdew1996} 

While both PBE0 and, above all, HSE06 have proved successful even when applied to solids,\cite{paier2006} it comes naturally to wonder if in this case a direct relationship between the exchange fraction $\alpha$ and some materials property can be established based on physical arguments. 
This possibility has been first investigated by Alkauskas~\emph{et al.},\cite{alkauskas2011} who identified static electronic screening as the essential ingredient determining the value of $\alpha$ in solids. This can be easily concluded by combining hybrid DFT and the static approximation to the $\mathit{GW}$ self-energy, i.e. Hedin's Coulomb-hole-plus-screened-exchange (COHSEX) approximation.\cite{hedin1965} The expression of the xc potential in the full-range hybrid functional scheme is
\begin{equation}
  v_{\text{xc}}(\bm{r}, \bm{r}') = \alpha v_{x}^{\text{ex}}(\bm{r}, \bm{r}') + (1 - \alpha) v_{x}^{\text{GGA}}(\bm{r})+v_{c}^{\text{GGA}}(\bm{r})
  \label{eq:hybrid-funct}
\end{equation}
where $v_{x}^{\text{ex}}$ is the nonlocal Hartree-Fock exact exchange potential evaluated using KS orbitals, and $v_{x}^{\text{GGA}}$ and $v_{c}^{\text{GGA}}$ are the local GGA exchange and correlation potentials, respectively. The COHSEX self-energy can be partitioned as follows
\begin{equation}
 \Sigma(\bm{r}, \bm{r}') = \Sigma_{\text{COH}}(\bm{r}, \bm{r}') + \Sigma_{\text{SEX}}(\bm{r}, \bm{r}')
 \label{eq:cohsex}
\end{equation}
where the local Coulomb-hole (COH) and nonlocal screened-exchange (SEX) potentials are given by

\begin{gather*}
 \Sigma_{\text{COH}}(\bm{r}, \bm{r}') = -\frac{1}{2} \delta (\bm{r} - \bm{r}')[v(\bm{r}, \bm{r}') - W(\bm{r}, \bm{r}')], \\
 \Sigma_{\text{SEX}}(\bm{r}, \bm{r}') = - \sum_{i=1}^{N_{\text{occ}}} \phi_i(\bm{r}) \phi_i(\bm{r}')  W(\bm{r}, \bm{r}').
\end{gather*}
The screened Coulomb potential is defined through
\begin{equation*}
 W(\bm{r}, \bm{r}') = \int d\bm{r}'' \epsilon^{-1}(\bm{r}, \bm{r}'') v(\bm{r}'', \bm{r}')
\end{equation*}
where $v$ is the bare Coulomb potential and $\epsilon^{-1}$ is the inverse of the microscopic dielectric matrix, the evaluation of which represents one of the most computationally demanding steps in $\mathit{GW}$ calculations. If microscopic contributions to dielectric screening are neglected and the inverse dielectric matrix is approximated as the inverse of the macroscopic electronic dielectric constant $\epsilon_{\infty}^{-1}$, then the screened potential becomes $ W(\bm{r}, \bm{r}') \approx \epsilon_{\infty}^{-1} v(\bm{r}, \bm{r}')$. Notice that the evaluation of the full microscopic dielectric matrix is now reduced to that of a ground-state, macroscopic, static quantity represented by the electronic dielectric constant $\epsilon_{\infty}$. It is clear that such simplification implies considerable reduction in the computational effort in comparison with $\mathit{GW}$. Finally, correspondence of spatially local and nonlocal contributions in Eqs.~\eqref{eq:hybrid-funct} and \eqref{eq:cohsex} leads to identifying 
\begin{equation}
\alpha=1/\epsilon_{\infty},
\label{eq:alpha}
\end{equation}
where $\epsilon_{\infty}$ is the static electronic dielectric constant of the material.\footnote{For a detailed derivation, see for example Ref.~\onlinecite{skone2014}} The corresponding hybrid xc functional is referred to as ``dielectric-dependent'' in the following.

Since the paper of Alkauskas~\emph{et al.} several works have appeared in the literature adopting this paradigmatic scheme and evaluating $\epsilon_{\infty}$ by means of different computational approaches.\cite{skone2014,marques2011,shimazaki2009, shimazaki2010,shimazaki2014,koller2013,refaely2013,moussa2012} Marques~\emph{et al.}\cite{marques2011} computed $\epsilon_{\infty}$ within the GGA Perdew-Burke-Ernzerhof\cite{perdew1996} approximation; alternatively, they suggested a relationship between $\alpha$ and a density-dependent estimator of the band gap, which they used for defining a material-dependent hybrid functional. Skone \emph{et al.}\cite{skone2014} recently suggested a self-consistent scheme for obtaining $\alpha$, by evaluating $\epsilon_{\infty}$ including local-field effects beyond the random-phase approximation (RPA). Conesa\cite{conesa2012} used instead the experimental dielectric constant. Other authors proposed adjusting $\alpha$ so as to reproduce the experimental band gap of materials,
\cite{alkauskas2008,broqvist2010,alkauskas2011} aiming at testing how the functional performs in predicting different electronic features, such as band edge positioning at semiconductor interfaces and defect energy levels in solids. Concerning the last point, Chen and Pasquarello\cite{chen2013} showed that good results are obtained if the parameter $\alpha$ is chosen in such a way that the modified hybrid functional matches the computed $\mathit{GW}$ band gap of the intrinsic material. 

In all previous work in which $\alpha$ is derived through Eq.~\eqref{eq:alpha}, attention has been focused on calculation of the electronic structure of semiconductors and insulators. In particular, the method has been shown to considerably improve over LDA, GGA and standard hybrid functionals concerning computation of the band gap for a wide class of materials. The authors of Refs.~\onlinecite{moussa2012, koller2013, skone2014} also computed optimized lattice parameters for a set of solids, showing that an accuracy comparable to that of standard hybrid functionals is achieved. 
What has not been assessed so far is if the method is also capable to provide good results in terms of the energetics and thermochemistry of semiconducting materials.

This is a crucial point to be addressed, both from a fundamental standpoint and in view of technological applications. For example, in the field of catalysis and photocatalysis, metal oxide semiconductors are widely employed.\cite{kudo2009} 
In a typical catalytic reaction, in which the semiconductor plays the role of the active catalyst, low-coordinated oxygen atoms at the semiconductor surface move into the reactant, leaving behind oxygen vacancies in the original material, which are subsequently filled by molecular oxygen present in the atmosphere. Thus, the semiconductor undergoes a change in stoichiometry for which an accurate description of the energetics is determinant in order to properly predict its performance as a catalyst.\cite{[For a recent review of the current challanges in the theoretical description of materials properties relevant to catalysis and photocatalysis see ] pacchioni2014} The formation of a vacancy also possibly strongly affects its electronic properties.  Accurately knowing the electronic band structure of the material is certainly paramount in the field of photocatalysis. For example, in semiconductor-based photocatalytic water 
splitting,\cite{currao2007} the band gap of the material has to exceed the redox potential of water ($\Delta E_{\text{redox}}^{\text{H$_2$O}}=1.23\,\text{eV}$), and in real devices should be no lower than 1.9 eV. 
Furthermore, the valence band maximum and conduction band minimum of the semiconductor have to lie higher than the water reduction potential and lower than the water oxidation potential, respectively, in order to be able to reduce or oxidize water and produce molecular hydrogen or oxygen. Wide gap metal oxide semiconductors have been identified as promising candidate materials for photocatalytic water splitting,\cite{kudo2009, vandekrol2008, ping2013review} as well as for catalysis, also due to their excellent optoelectronic properties.

In this scenario, an assessment of the performance of the dielectric-dependent hybrid approach in predicting electronic and structural properties of oxide semiconductors, as well as the energetics pertaining different kinds of transformations in such materials is certainly in order. We have selected a set of wide gap oxide semiconductors with current or potential technological impact, e.g. in the fields of catalysis, photocatalysis, and photovoltaics. As exemplified above, several issues concerning both the electronic structure and the energetics are to be tackled in these materials in order to properly conceive practical applications. To this purpose, it would be extremely useful to rely on a single theoretical framework in which all such properties can be accurately predicted from first-principles in a unified scheme. We have benchmarked the dielectric-dependent hybrid functional approach as a candidate method in this respect, allowing to reliably compute both band structures and total energies at 
reasonable computational cost.

In the present work we systematically investigate band gaps and ground-state properties for the following set of oxide semiconductors in their bulk structures: ZnO, TiO$_2$, ZrO$_2$, WO$_3$; all are interesting materials for several applications in the energy field, as discussed above. Results are also reported for MgO, which is studied as a simple test material although being an insulator with a gap as large as 8 eV. DFT calculations have been performed using different local/semilocal and hybrid functionals, comparing plane waves and localized basis sets within the linear combination of atomic orbitals (LCAO) method. $\mathit{GW}$ band gaps for selected polymorphs of the above oxides have also been computed for comparison. Then the relative stability of several polymorphs is investigated at different levels of theory. Finally, the case of the chemical reduction of TiO$_2$ to Ti$_2$O$_3$, involving a change in oxide stoichiometry, is analyzed; it represents a prototypical 
chemical process central to catalysis. The last two points require investigating, although from different perspectives, the common issue of the energetics. For all of the above properties, we report a systematic comparison of at least one approach belonging to the classes of local/semilocal, standard hybrid, and dielectric-dependent hybrid functionals.

\section{Computational methods}

\subsection{Plane waves}

Plane wave (PW) DFT calculations were performed using the \textsc{quantum espresso} package.\cite{QE-2009} A norm-conserving Troullier-Martins pseudopotential\cite{troullier1991} with $2s$ and $2p$ electrons in the valence was used for oxygen, while for metals, norm-conserving Hartwigsen-Goedecker-Hutter pseudopotentials\cite{goedecker1996, hartwigsen1998} explicitly including semicore electrons in the valence were employed.
A careful treatment of core-valence partitioning in metal atoms has been proved to be determinant for obtaining reliable $\mathit{GW}$ corrections,\cite{marini2001} especially in II-IV semiconductors, as analyzed in detail in Ref.~\onlinecite{rohlfing1995}. It turns out that explicit treatment of semicore electrons of the metal atom in oxides is always needed, as we experienced in the case of ZnO (see discussion in Section~\ref{subsec:bandgap}). Since our PW DFT calculations served as starting point for subsequent $\mathit{GW}$ 
calculations, our choice of the pseudopotentials was guided 
by the above 
considerations. This means that $(n-1)s$ and $(n-1)p$ electrons of the metal atom were 
always explicitly 
treated in the valence, where $n$ 
is the main quantum number of the outermost occupied electronic shell. 
Pseudopotentials were tested by checking that computed optimized lattice parameters and electronic band structures were 
in agreement with the reference 
literature, as well as that the former showed the expected trends at the LDA and GGA levels of theory when compared to experiment (see the Supplemental Material).\footnote{See the Supplemental Material for the experimental lattice parameters, Brillouin-zone samplings, plane-wave kinetic energy cutoffs and various cutoffs defined in $\textit{GW}$ calculations; results of geometry optimizations not presented in the main text are also reported.}

Including semicore electrons in the self-consistent calculation of charge density clearly requires using a large number of PW components in the expansion of the KS orbitals, as semicore electrons are tightly bound to the nucleus. This dramatically affects the resulting computational cost, especially in hybrid functional calculations, in which evaluation of the nonlocal exact exchange using a PW basis set under periodic boundary conditions is particularly demanding.

Brillouin zone (BZ) sampling is also critical for obtaining well-converged results. We chose Monkhorst-Pack\cite{monkhorst1976} (MP) $k$-point grids in such a way that estimated errors in computed band gaps and total energies were within 20~meV. In hybrid calculations, evaluation of exact exchange energy in PW basis requires the definition of an additional grid of $q$-points, which must be a symmetry-preserving subset of the $k$-point grid. The choice of both $k$- and $q$-point grids determines the accuracy of the computed quantities, while strongly affecting the computational cost (see the Supplemental Material).

Full structural optimizations were carried out only at the LDA and GGA levels, as optimizations with hybrid functionals are very expensive with \textsc{quantum espresso}. A quasi-Newton Broyden–-Fletcher-–Goldfarb–-Shanno (BFGS) scheme was adopted for the search of the energy minimum. Convergence thresholds on atomic forces and pressure were set to $10^{-3}$~a.u. and $0.5$~kbar, respectively.

\subsection{Localized basis sets}

The LCAO method, as implemented in the \mbox{\textsc{crystal09}} package,\cite{dovesi2005} was employed for DFT calculations using localized basis sets. All-electron calculations were performed for MgO, ZnO and TiO$_2$, while for ZrO$_2$ and WO$_3$ the inner electrons of Zr and W atoms were described through effective core pseudopotentials (ECPs). In the \textsc{crystal09} code, KS orbitals are expressed in terms of localized atomic-like functions which are in turn expanded in a linear combination of Gaussian-type orbitals (GTOs) with fixed coefficients. Expansion coefficients and exponents of GTOs define a basis set for a given atom. 

In the present work, we adopted all-electron and valence basis sets which were previously tested in analogous solid state calculations. The following Gaussian-type all-electron basis sets were employed in our calculations: \mbox{8-511(d1)} for Mg from Ref.~\onlinecite{valenzano2006}, \mbox{pob-TZVP} for Zn from Ref.~\onlinecite{peintinger2013}, \mbox{8-4611(d41)} for Ti from Ref.~\onlinecite{wilson1998}. For O the \mbox{8-411(d1)} basis set of Ref.~\onlinecite{ruiz2003} was used, except in the case of ZrO$_2$, in which the basis set from Ref.~\onlinecite{noel2001} was adopted, following the choice of a previous investigation on this material.\cite{gallino2011} Small-core Hay-Wadt ECPs\cite{hay1985} were employed for the heavy Zr and W atoms. For Zr only the $4d$ and $5sp$ electrons were explicitly treated, using the \mbox{311(d31)} valence electron basis set of Ref.~\onlinecite{bredow2004}. For W, electrons belonging to shells $5p$, $5d$, $6sp$ were included in
the valence, and a modified Hay-Wadt \mbox{double-$\zeta$} basis set\cite{wang2011} was employed to describe them. 

In the analysis of the chemical reduction of TiO$_2$ to Ti$_2$O$_3$, for both Ti and O atoms, we adopted different basis sets than mentioned above, following previous Hartree-Fock and DFT+U studies\cite{catti1997,hu2011} on the same materials. In molecular calculations we chose basis sets for oxygen\cite{ruiz2003} and hydrogen\cite{gatti1994} such that experimental atomization energies for H$_2$, O$_2$ and H$_2$O molecules were well-reproduced at the PBE0 level.

As a high accuracy is needed in studying the energetics, BZ samplings were performed so as to ensure convergence within 1~meV on total energies for the different polymorphs of the materials under investigation. This corresponds to MP grids including a number of $k$-points in the irreducible BZ at least equal to what we used in PW calculations (see the Supplemental Material). The lower cost of LCAO calculations made this choice affordable. 

Different cutoff thresholds for the evaluation of Coulomb and exchange integrals written in terms of GTOs were set to their standard values in \textsc{crystal09}.\footnote{The following thresholds were used: $10^{-7}$ for Coulomb overlap tolerance, $10^{-7}$ for Coulomb penetration tolerance, $10^{-7}$ for exchange overlap tolerance, $10^{-7}$ for exchange pseudo-overlap in direct space, $10^{-14}$ for exchange pseudo-overlap in reciprocal space.} The self-consistent field was 
considered converged when total energy difference between two subsequent cycles became lower than $10^{-6}$~a.u. We kept these thresholds fixed throughout all the calculations for a more meaningful comparison of the computed total energies.

Geometry optimizations were performed at both local/semilocal and hybrid functional level by fully optimizing both lattice parameters and atomic coordinates. A quasi-Newton algorithm with a BFGS Hessian updating scheme was adopted for the search of the energy minimum. Geometry optimizations were terminated when the maximum component and root-mean-square (rms) of energy gradients became lower than $0.00045$ and $0.00030$~a.u., respectively, and the maximum and rms atomic displacements were below $0.00180$ and $0.00120$~a.u., respectively. 

\subsection{$\mathit{GW}$ calculations}

$\mathit{GW}$ calculations were performed using the \textsc{berkeleygw} code.\cite{deslippe2012, hybertsen1986} The non self-consistent $G_0W_0$ scheme was adopted, in which both the electronic Green's function and the screened Coulomb potential were constructed starting from DFT-KS eigenfunctions $\psi_{n\bm{k}}^{\text{KS}}$ and eigenvalues $E_{n\bm{k}}^{\text{KS}}$ computed at the PBE level of theory ($G_0W_0$@PBE). Quasiparticle (QP) energies corresponding to band index $n$ are evaluated on top of KS eigenvalues using the first-order perturbation formula
\begin{equation}
  E_{n\bm{k}}^{\text{QP}} = E_{n\bm{k}}^{\text{KS}} + \left<\psi_{n\bm{k}}^{\text{KS}}\right|\Sigma(E_{n\bm{k}}^{\text{QP}})-v_{xc}^{\text{PBE}} \left|\psi_{n\bm{k}}^{\text{KS}}\right>.
  \label{eq:qp-energy}
\end{equation}
This implicit equation is solved by using Newton's method to linearize the energy-dependent self-energy $\Sigma(E)$ around the zero-order QP energy\cite{deslippe2012}

\begin{equation*}
  E_{n\bm{k}}^{0} = E_{n\bm{k}}^{\text{KS}} + \left<\psi_{n\bm{k}}^{\text{KS}}\right|\Sigma(E_{n\bm{k}}^{\text{KS}})-v_{xc}^{\text{PBE}} \left|\psi_{n\bm{k}}^{\text{KS}}\right>,
\end{equation*}
finally yielding

\begin{equation*}
  E_{n\bm{k}}^{\text{QP}} \simeq E_{n\bm{k}}^{0} + \frac{d\Sigma/dE}{1-d\Sigma/dE}\left(E_{n\bm{k}}^{0} - E_{n\bm{k}}^{\text{KS}} \right).
\end{equation*}

No further iteration was carried out on either the Green's function or the self-energy. In fact, whether a self-consistent $\mathit{GW}$ approach really improves computed band gaps over $G_0W_0$ is still matter of debate.\cite{shishkin2007}  DFT calculations were performed at the experimental geometry using \mbox{\textsc{quantum espresso}}.
The dielectric function was evaluated at the RPA level, the frequency dependence being accounted for through the plasmon-pole approximation of Hybertsen and Louie.\cite{hybertsen1986}

In the \textsc{berkeleygw} code, all the quantities of interest are expressed in a PW basis set, evaluating them in reciprocal space. Summations over a large number of empty states are required for obtaining well-converged polarizability and Coulomb hole self-energy. This calls for extensive convergence studies on various cutoff parameters entering calculations. The size of the static dielectric matrix in reciprocal space is controlled through the cutoff energy $E_{\text{cut}}^{\text{eps}}$. Additionally, the number of empty states in the expressions of the static RPA polarizability $\chi$, and of the Coulomb hole (CH) contribution to dynamical self-energy, $\Sigma_{\text{CH}}$, should be determined. Polarizability typically converges more rapidly than self-energy, for which we found that no less than 900 empty states were needed in our calculations (see the Supplemental Material). Finally, the number of terms in the PW expansion of the xc potential, $v_{xc}$, in Eq.~\eqref{eq:qp-energy} is determined 
by the cutoff energy $E_{\text{cut}}^{\text{xc}}$. Care should be taken when performing convergence tests, as not all cutoff parameters can be converged independently of each other. In fact, the size of the dielectric matrix in reciprocal space is not independent of the number of empty states included  in $\Sigma_{\text{CH}}$. In some previous work, this tricky behavior has not been considered with the due attention, leading to false convergence issues, as pointed out by Louie and coworkers\cite{shih2010} for the particularly critical case of ZnO. 

In our calculations, careful convergence studies were carried out for all the materials, ensuring that accuracy within 50~meV on QP band gaps was reached. To this purpose, we closely followed the procedure described in Ref.~\onlinecite{malone2013}, to which we refer for further details concerning the implementation of the $\mathit{GW}$ scheme in \textsc{berkeleygw}, as well as for a description of the steps to be followed for assessing the convergence of calculated quantities. In Table~S2 of the Supplemental Material  cutoff 
parameters used in our calculations are reported. As they are not expected to depend strongly on the BZ sampling,\footnote{See the Supplemental Material of Ref.~\onlinecite{malone2013}} we estimated them using a minimal 1x1x1 grid. For the fully-converged $\mathit{GW}$ calculations, BZ samplings were performed as in DFT PW calculations.

\subsection{DFT xc functionals}
\label{sec:comp-det}

In the first part of the work, we present our computed band gaps using different local/semilocal and hybrid xc functionals, comparing PW and LCAO approaches. We also compare band gaps computed using both dielectric-depended hybrid functional and $\mathit{GW}$ approaches.

In our local/semilocal calculations we used the Perdew-Zunger expression for the correlation energy\cite{perdew1981} for the LDA, and the Perdew-Burke-Ernzerhof (PBE) parametrization\cite{perdew1996} for the GGA. Hybrid functionals considered are the full-range PBE0\cite{perdew1996-pbe0} with exchange fraction $\alpha=1/4$ and the screened-exchange HSE06\cite{heyd2003, heyd2006} with $\alpha=1/4$ and range of the screened exchange interaction set to its standard value $\omega=0.11$~a.u.$^{-1}$.\footnote{The \textsc{crystal14} code was used for HSE06 calculations, as screened-exchange hybrid functionals are not implemented in \textsc{crystal09}.} We also tested the B3LYP hybrid functional,\cite{becke1993b,stephens1994} which combines a fraction $\alpha = 1/5$ of 
Hartree-Fock exact exchange with the GGA Becke-Lee-Yang-Parr (BLYP) xc functional.\cite{lee1988} 

In hybrid calculations with \textsc{quantum espresso}, we chose pseudopotentials generated at 
the closest GGA level, i.e. PBE for PBE0 and HSE06, and 
BLYP for B3LYP. We are 
not aware, to date, of any publicly available packages for pseudopotential generation with hybrid functionals. Use of a Hartree-Fock instead of a GGA pseudopotential in a previous study on silicon\cite{[See the Supplemental Material of ] jain2011} showed that little effect should be observed on the computed band gap.\cite{[See ][ for a thorough discussion on the transferability of \emph{ab initio} pseudopotentials outside the theoretical framework used to generate them.] focher1991} We obtained all pseudopotentials from the \textsc{quantum espresso} pseudopotential library,\cite{qe-pp} except for the Zn atom, for which the pseudopotential was converted from the \texttt{cpmd} format.\cite{goedecker1996, hartwigsen1998, krack2005,* krack-pp}

We constructed dielectric-dependent hybrid functionals by defining the exchange fraction according to Eq.~\eqref{eq:alpha}. Electronic dielectric constants were computed within the couple-perturbed Kohn-Sham (CPKS) method\cite{ferrero2008a, ferrero2008b, ferrero2008c} as implemented in the \textsc{crystal09} code. This method exploits first-order perturbation theory to calculate the polarizability, and fully includes local-field effects, as discussed in Ref.~\onlinecite{skone2014}. We evaluated $\epsilon_{\infty}$ both at the PBE and PBE0 levels, thus defining the PBE0$\alpha_{\text{PBE}}$ and PBE0$\alpha_{\text{PBE0}}$ dielectric-dependent functionals. The inclusion of some degree of self-consistency in the calculation of $\alpha$ was also tested in certain cases.

\section{Results and discussion}

\subsection{Band gap: local/semilocal, hybrid functionals and $\mathit{GW}$}
\label{subsec:bandgap}

In this Section, the following materials are investigated: cubic (c) rocksalt MgO, wurtzite (wz) ZnO, anatase (a) TiO$_2$, tetragonal (t) WO$_3$, room-temperature $\gamma$-monoclinic ($\gamma$-m) WO$_3$. We computed band gaps within DFT (with both PW and LCAO schemes) using different functionals, as well as within $\mathit{GW}$; results are reported and compared in Table~\ref{tab1}, along with selected experimental results. For a more meaningful comparison of the different computational approaches, and due to the prohibitive cost of geometry optimizations using hybrid functionals with \mbox{\textsc{quantum espresso}}, calculations in this Section were performed at the experimental geometry (see the Supplemental Material).

We emphasize that comparison between theoretical calculations and experimental data is typically far from straightforward when it comes to spectroscopic properties of semiconductors and insulators. From the experimental point of view, different techniques give access to different physical observables. In (direct and inverse) photoemission experiments, electrons are removed from or added to the solid (charged excitations), probing hence the energy difference between the $N$ and the $N \pm 1$ particles systems. From the theoretical point of view, such total energy differences are described by the poles of the one-particle Green's function $G$, commonly named the QP energies. Thus, the measured QP energies or their difference (the fundamental gap) can be put in direct comparison with values obtained by solving the QP eigenvalue problem,\cite{hybertsen1986} after devising an approximation to the self-energy operator $\Sigma$, for example within the $\mathit{GW}$ scheme, or by approximating it with a DFT xc 
potential (in which case the QP energies effectively reduce to the KS eigenvalues).

In contrast, optical measurements involve neutral excitations, which can be seen as creation of electron-hole pairs. The binding energy of electron-hole pairs is hence inherently included. The theoretical description of neutral excitations requires in principle working with the two-particle (electron-hole) Green's function, and such an analysis lies beyond the scope of the present work. Nonetheless, by knowing or guessing the exciton binding energy, optical measurements can be useful for extrapolating QP gaps when direct and inverse photoemission measurements are not available for the material at hand. This situation is not uncommon, as discussed in the following for our investigated materials.
 
As a second source of possible disagreement with experiment, we point out that our calculations, being performed at frozen ion positions, do not take into account electron-phonon interaction, which generally leads to renormalization of the band gap, reducing it with respect to the zero-temperature QP gap.\cite{cardona2005, giustino2010}

Given the above considerations, and before discussing the results of our calculations, it is worthwhile to give an overview of the experimental situation.  

\subsubsection{Experimental data}

The fundamental gap reported for MgO was extrapolated from thermoreflectance studies\cite{whited1973} following estimation of the exciton binding energy. 

In the case of ZnO, different optical experiments\cite{liang1968, reynolds1999} agree on the value of our reported QP gap, which was extracted from optical experiments by detailed analysis of the excitonic levels. 

For anatase TiO$_2$, the situation is more complicated, since, to our knowledge, no experimental data are available for the fundamental gap. In Ref.~\onlinecite{tang1995}, analysis of temperature dependence of the absorption coefficient allowed to estimate the indirect optical gap to be $3.42$ eV. For rutile TiO$_2$, measurements of both the QP and the optical band gap are available. An optical band gap of $\sim 3.0$~eV has been reported in several investigations,\cite{pascual1977,tang1995} while combined photoemission and inverse photoemission experiments yielded for the direct fundamental band gap the values of $3.3\pm 0.5$~eV (Ref.~\onlinecite{tezuka1994}) and $3.6 \pm 0.2$~eV (Ref.~\onlinecite{rangan2010}). Assuming a similar relationship between fundamental and optical gap holds for anatase, we could tentatively estimate its QP gap to be in the range $3.7$--$4.0$ eV, as also suggested in Refs.~\onlinecite{kang2010, landmann2012}. 

The only available data for the pure phases of ZrO$_2$ are from optical absorption measurements,\cite{french1994} from which a direct optical gap of $5.78$~eV for the tetragonal phase was inferred. Photoemission studies yielded smaller measured gaps of $5.68$~eV (Ref.~\onlinecite{sayan2004}) and $5.5$~eV (Ref.~\onlinecite{bersch2008}); however these values cannot be attributed to any of the three ZrO$_2$ phases, as the probed samples were amorphous or polyscrystalline. 

For WO$_3$, several ultraviolet direct/inverse photoemission studies\cite{meyer2010,kroeger2009,weinhardt2008} have reported a band gap in the range $3.28$--$3.39$~eV. Instead, absorption measurements\cite{koffyberg1979,hodes1976,
bringans1981} have shown large data dispersion, with an optical band gap spanning the range $2.6$--$3.2$~eV.

\begin{table*}[tb]
\caption{\label{tab1} Fundamental band gaps (in eV) of the various oxides computed within local/semilocal, standard hybrid, and $\mathit{G_0W_0}$ schemes, comparing PW and LCAO calculations. All calculations were performed at the experimental geometry. When the fundamental gap is indirect, direct band gaps are reported in parenthesis. Estimated errors are 20~meV and 50~meV for the DFT and $GW$ results, respectively.}
\begin{ruledtabular}
\begin{tabular}{lllccccccc} 
 & Type & Method & LDA & PBE & PBE0 & HSE06 & B3LYP & $G_0W_0$@PBE & Expt.\\
\hline
\multirow{2}{*}{MgO} & \multirow{2}{*}{rs} & PW & 4.63 & 4.79 & 7.18 & 6.44 & 7.04 & 7.88 & \multirow{2}{*}{7.83\footnote{Thermoreflectance at 85 K, Ref.~\onlinecite{whited1973} }}\\
& & LCAO & 4.86 & 4.93 & 7.35 & 6.67 & 7.07 \\
\\
\multirow{2}{*}{ZnO} & \multirow{2}{*}{wz} & PW & 0.78 & 0.81 & 3.18 & 2.46 & 2.90 & 3.06 & \multirow{2}{*}{3.44\footnote{Transmission spectroscopy at 4.2 K, Ref.~\onlinecite{liang1968}; photoluminiscence at 2 K, Ref.~\onlinecite{reynolds1999}}}\\
 & & LCAO & 0.92 & 1.02 & 3.35 & 2.75 & 3.05 \\
\\
\multirow{2}{*}{TiO$_2$} & \multirow{2}{*}{a} & PW & 2.03 (2.16) & 2.13 (2.26) & 4.40 (4.47) & 3.67 (3.74) & 3.94 (4.03) & 3.73 (3.81) & \multirow{2}{*}{3.42\footnote{Transmission spectroscopy, Ref.~\onlinecite{tang1995}}} \\
& & LCAO & 2.09 (2.18) & 2.18 (2.28) & 4.23 (4.33) & 3.59 (3.70) & 3.79 (3.89) \\
\\
\multirow{2}{*}{ZrO$_2$} & \multirow{2}{*}{t} & PW & 4.01 (4.18) & 4.12 (4.29) & 6.52 (6.69) & 5.78 (5.95) & 6.09 (6.25) & 5.87 (6.06) & \multirow{2}{*}{5.78\footnote{Reflectance spectroscopy (VUV) in Y-doped ZrO$_2$, Ref.~\onlinecite{french1994}}} \\
& & LCAO & 3.94 (4.09) & 4.00 (4.16) & 6.33 (6.45) & 5.61 (5.75) & 5.90 (6.02) \\
\\
\multirow{2}{*}{WO$_3$} & \multirow{2}{*}{$\gamma$-m} & PW & 1.84 & 1.94 & 3.90 & 3.16 & 3.53 & 3.34 & \multirow{2}{*}{3.38\footnote{Photoemission (UPS-IPES), Ref.~\onlinecite{meyer2010}}} \\
& & LCAO & 1.86 & 1.91 & 3.74 & 3.09 & 3.39 \\
\end{tabular}
\end{ruledtabular}
\end{table*}

\subsubsection{DFT LDA, GGA and standard hybrids: comparison between PW and LCAO}

Results of our calculations confirm that, as expected, hybrid functionals provide substantial improvement over well-known LDA/GGA gap underestimation. In particular, band gaps computed at the HSE06 level are in good agreement with experiment for a-TiO$_2$, t-ZrO$_2$ and $\gamma$-m-WO$_3$, as also confirmed by previous calculations on titanium\cite{landmann2012,deak2011,janotti2010} and tungsten\cite{wang2011} oxides. For the same materials, we found that PBE0 overestimates the band gap with respect to experiment, while B3LYP generally gives better results; this behavior is in agreement with previous investigations.\cite{gallino2011,wang2011,beltran2001,zhang2005,calatayud2001} Quantitative disagreement with the literature, when present, may be attributed to different choices of geometry (optimized vs. experimental, the latter being considered in the present part of the work). Independently of the adopted theoretical treatment, we obtained an indirect gap for a-TiO$_2$ and t-ZrO$_2$ and a direct 
gap for the 
other 
materials. Both values are reported in Table~\ref{tab1}. 

In the case of MgO and ZnO, we observe that hybrid functionals perform differently from what previously discussed. MgO is an insulator with a wide band gap of nearly 8~eV; its computed gap, even using hybrid functionals, is smaller than the experimental value. In the case of ZnO, due to the strong gap underestimation at the LDA/GGA level, and contrary to what we found for the other semiconductors, PBE0 seems to perform well, similarly to B3LYP, while HSE06 is not sufficient to correct the PBE underestimation. Recent studies using hybrid functionals\cite{skone2014,marques2011,tosoni2012,wan2011,wrobel2009} confirm the observed trends for MgO and ZnO.

We finally comment on the comparison between results obtained within PW and LCAO schemes. From Table~\ref{tab1}, it is found that discrepancies in computed band gaps range from $\sim 0.05$ to $\sim 0.30$~eV, most of them being within $0.20$~eV, depending on the material and the functional used. The closest correspondence is found at the LDA/GGA level, while with hybrid functionals differences become more substantial. Different results are the consequence of different kinds of approximations characterizing the two computational schemes. Generation of reliable pseudopotentials on the one hand and construction of accurate basis sets for solid state calculations on the other, are certainly two of the main critical points in this respect. For example, in the case of ZnO, using a Zn pseudopotential with only $3d$ and $4s$ electrons in valence, and adopting a non optimal basis set for Zn in LCAO 
calculations, resulted in computed gaps differing by as much as $\sim 
0.5$~eV 
within LDA/GGA and  $\sim 1$~eV at the hybrid functional level. Adoption of a better pseudopotential and basis set indeed partially corrected this discrepancy. Inclusion of semicore electrons in Zn has also been found to be crucial for obtaining reliable $\mathit{G_0W_0}$ corrections over the PBE gap,\footnote{Our computed $\mathit{G_0W_0}$ gap without explicitly treating semicore $3s$ and $3p$ electrons is $\sim 1$~eV.} similarly to what was reported by Gori~\emph{et al.}\cite{gori2010} In conclusion, we believe that, given the above discussion, the agreement between our computed PW and LCAO band gaps should be deemed satisfactory. 
\begin{table*}[tb]
\caption{\label{tab2} Band gap energy (in eV) computed using dielectric-dependent hybrid functionals defined through the exchange fractions reported in Table~\ref{tab3}. Comparison with standard PBE0 and $\mathit{G_0W_0}$ is provided. Only the direct gap is reported for ZrO$_2$, since the measured gap is extrapolated via a model for a direct gap. MAE and MARE are the mean absolute and mean absolute relative error, respectively, with respect to the computed $\mathit{G_0W_0}$ band gap. All DFT calculations were performed within the LCAO scheme at the experimental geometry. Estimated errors are 20~meV and 50~meV for the DFT and $GW$ results, respectively.}
\begin{ruledtabular}
\begin{tabular}{lldddcc}
& Type & \multicolumn{1}{c}{PBE0$\alpha_{\text{PBE}}$} & \multicolumn{1}{c}{PBE0$\alpha_{\text{PBE0}}$} & \multicolumn{1}{c}{PBE0} & $\mathit{G_0W_0}$@PBE & Expt. \\
\hline
MgO & rs & 8.06 & 8.33 & 7.38 & 7.88 & 7.83 \\
ZnO & wz & 3.18 & 3.94 & 3.35 & 3.06 & 3.44 \\
TiO$_2$ & a & 3.38 & 3.72 & 4.24 & 3.73 & 3.42 \\
ZrO$_2$ & t & 5.81 & 6.12 & 6.45 & 6.06 & 5.78 \\
WO$_3$ & $\gamma$-m & 3.23 & 3.50 & 3.74 & 3.34 & 3.38 \\
\\
MAE (eV) &  & 0.20 & 0.31 & 0.42 &  &  \\
MARE (\%) & & 4.6 & 8.1 & 9.6 &  &  \\
\end{tabular}
\end{ruledtabular}
\end{table*}

 \begin{table}[tb]
\caption{\label{tab3} Electronic dielectric constant $\epsilon_{\infty}$ (estimated error within 0.02) and corresponding exchange fraction $\alpha$ (\%) evaluated at the PBE and PBE0 levels. Calculations were performed within the LCAO scheme at the experimental geometry. Measured dielectric constants are provided for comparison.}
\begin{ruledtabular}
\begin{tabular}{llcccccd}
&  \multirow{2}{*}{Type} & \multicolumn{2}{c}{PBE} & \multicolumn{2}{c}{PBE0} 
& \multicolumn{1}{c}{Expt.\footnote{Measured static electronic dielectric constants from the following references: MgO, Ref.~\onlinecite{lide2010}; ZnO, Ref.~\onlinecite{ashkenov2003}; TiO$_2$, Ref.~\onlinecite{wemple1977}; ZrO$_2$, Ref.~\onlinecite{french1994}, WO$_3$, Ref.~\onlinecite{hutchins2006}}} \\
 \cline{3-4} \cline{5-6} \cline{7-7}
 & & $\epsilon_{\infty}$ & $\alpha$ & $\epsilon_{\infty}$ & $\alpha$ & \multicolumn{1}{c}{$\epsilon_{\infty}$} \\
\hline
MgO & rs & 3.10 & 32.2 & 2.87 & 34.9 & 2.96 \\
ZnO & wz & 4.31 & 23.2 & 3.24 & 30.8 & 3.74  \\
TiO$_2$ & a & 6.52 & 15.3 & 5.16 & 19.4 & 5.62 \\
ZrO$_2$ & t & 5.45 & 18.3 & 4.62 & 21.6 & 4.9 \\
WO$_3$ & $\gamma$-m & 5.43 & 18.4 & 4.56 & 21.9 & 4.81 \\
\end{tabular}
\end{ruledtabular}
\end{table}

\subsubsection{Many-body perturbation theory: $\mathit{G_0W_0}$}

Table~\ref{tab1} reports $\mathit{G_0W_0}$ band gaps computed on top of PBE calculations. We notice that generally good agreement is obtained with experiment for all the materials. Our results are in line with previous investigations, although numerical values may differ as a consequence of the different computational setup adopted, e.g. the starting DFT approximation to QP energies and wavefunctions, and the way the frequency dependence of the dielectric function is accounted for.\cite{kang2010} 

For MgO, our computed gap of $7.88$~eV is very close to experiment and to the results of the $\mathit{G_0W_0}$ investigation of Fuchs~\emph{et al.} adoping a HSE03 starting point.\cite{fuchs2007} Calculations starting from DFT-PBE using the projected augmented wave (PAW) method gave smaller QP gaps of $7.25$~eV (Ref.~\onlinecite{shishkin2007}) and $7.41$~eV (Ref.~\onlinecite{chen2012}). 

For TiO$_2$, our obtained indirect QP gap of $3.73$~eV is identical to the full-frequency result of Landmann~\emph{et 
al.},\cite{landmann2012} and in close agreement with other investigations.\cite{thulin2008,kang2010,chiodo2010,patrick2012} 

Quasiparticle band structure calculations of ZrO$_2$ are scarce. The only $\mathit{G_0W_0}$ studies we are aware of reported a fundamental band gap of $6.40$~eV (Ref.~\onlinecite{kralik1998}) and $5.56$~eV (Ref.~\onlinecite{jiang2010}) for tetragonal zirconia. Our result of $6.06$~eV for the direct gap is in between these two values, and compatible with the measured optical gap of $5.87$~eV. 

As for $\gamma$-m-WO$_3$, recent studies of Galli and coworkers yielded QP gaps of 3.26~eV (Ref.~\onlinecite{ping2013}) and 3.30~eV (Ref.~\onlinecite{ping2014}), which are very close to our result, and in excellent agreement with the measured photoemission gap. 

Finally, ZnO stands out as a particularly critical case, for which considerable disagreement is found in the theoretical literature concerning its QP properties. Several studies have yielded QP band gaps ranging from $\sim 2.1$~eV (Refs.~\onlinecite{shishkin2007,fuchs2007}) to $3.4$~eV obtained by Louie and coworkers
\cite{shih2010} within $\mathit{G_0W_0}$@LDA. These authors suggested that smaller computed gaps may be the consequence of false convergence behavior of the dielectric function and self-energy. Stankovski~\emph{et al.}\cite{stankovski2011} studied the effect of different plasmon-pole models on the band gap, while Friedrich~\emph{et al.}\cite{friedrich2011} carried out $\mathit{G_0W_0}$ on top of all-electron DFT calculations to investigate the effect of the pseudopotential approximation. At present there is no general consensus onto which is the most correct approach to describe the QP band structure of this material. Our calculations gave a gap significantly smaller than the experiment, which is in line with several previous investigations within the $\mathit{G_0W_0}$@LDA/GGA approach.
 
\subsubsection{DFT dielectric-dependent hybrid functionals} 

We turn to study the performance of different kinds of dielectric-dependent hybrid functionals. The fraction of exact exchange is defined through $\alpha=1/\epsilon_{\infty}$; the static electronic dielectric tensor was computed within both PBE and PBE0; $\epsilon_{\infty}$ is defined as the average of the diagonal elements of the diagonalized dielectric tensor written in real space.
 
Table~\ref{tab3} reports the computed dielectric constants along with the corresponding exchange fractions $\alpha_{\text{PBE}}$ and $\alpha_{\text{PBE0}}$, defining the modified PBE0 functionals, PBE0$\alpha_{\text{PBE}}$ and PBE0$\alpha_{\text{PBE0}}$. We notice that, as expected, PBE and PBE0 systematically overestimates and underestimates, respectively, the dielectric constant, as a consequence of the opposite tendency concerning estimation of the band gap. 

Table~\ref{tab2} shows that the dielectric-dependent hybrid functional approach remarkably improves the computed gap over standard fixed-$\alpha$ PBE0. Agreement with experiment and $\mathit{G_0W_0}$ calculations is overall satisfactory with both modified hybrids, with the exception of ZnO, for which standard PBE0 already performs well. Since reliable experimental photoemission gaps are not available for all the materials, mean errors in Table~\ref{tab2} are defined with respect to the computed $\mathit{G_0W_0}$ gaps. Error analysis shows that the PBE0$\alpha_{\text{PBE}}$ functional achieves the closest agreement with $\mathit{G_0W_0}$ for the class of materials under investigation. We thus took it as our preferred starting dielectric-dependent hybrid approach in the rest of the work.

\subsection{Oxide polymorphs: band gaps, crystal structures and phase stability}

We now turn to investigation of several other structural modifications of our set of oxide semiconductors. This includes analysis of the their relative phase stability, electronic and dielectric properties, and equilibrium crystal structure, by using different theoretical methods. 

Phase stability is evaluated on the basis of ground-state total energy calculations, thus not taking into account thermal and entropic contributions. In particular, the thermal contribution mainly comes from lattice vibrations (phonons), and may be relevant in temperature-induced structural phase transitions. Here we argue that an analysis of the only zero-temperature electronic contribution is still able to provide a qualitative picture on the stability issue. In fact, although in this case quantitative comparison with thermochemical data may be of limited significance, it can be interesting to assess whether theory is able to correctly reproduce the experimentally found phase stability sequence (at ambient pressure) for a given material. 

\begin{table}[tb]
\caption{\label{tab4} Electronic dielectric constant $\epsilon_{\infty}$ (estimated error within 0.02) and corresponding exchange fraction $\alpha=1/\epsilon_{\infty}$ (\%) evaluated within PBE and dielectric-dependent PBE0 (PBE0$\alpha_{\text{PBE}}$ and PBE0$\alpha_{\text{PBE}}^{\text{(1)}}$) for various polymorphs of the studied oxides. Calculations were performed within the LCAO scheme at the optimized geometry.}
\begin{ruledtabular}
\begin{tabular}{llcccccc}
 & \multirow{2}{*}{Type} & \multicolumn{2}{c}{PBE} & \multicolumn{2}{c}{PBE0$\alpha_{\text{PBE}}$} & \multicolumn{2}{c}{PBE0$\alpha_{\text{PBE}}^{\text{(1)}}$} \\
 \cline{3-4} \cline{5-6} \cline{7-8}
 & & $\epsilon_{\infty}$ & $\alpha$ & $\epsilon_{\infty}$ & $\alpha$ & $\epsilon_{\infty}$ & $\alpha$ \\
\hline
 MgO & rs & 3.13 & 31.9 & 2.80 & 35.7 & 2.78 & 36.0 \\
\\
\multirow{2}{*}{ZnO} & wz & 4.30 & 23.2 & 3.28 & 30.8 & 3.14 & 31.8 \\
                     & zb & 4.67 & 21.4 & 3.37 & 29.6 & 3.19 & 31.3 \\
\\
\multirow{3}{*}{TiO$_2$} & r & 7.98 & 12.5 & 6.76 & 14.8 & 6.59 & 15.2 \\
                         & a & 6.58 & 15.2 & 5.59 & 17.9 & 5.45 & 18.3 \\
                         & b & 6.95 & 14.4 & 5.99 & 16.7 & 5.87 & 17.0 \\
\\
\multirow{3}{*}{ZrO$_2$} & m & 5.29 & 18.9 & 4.62 & 21.6 & 4.55 & 22.0 \\
                         & t & 5.56 & 18.0 & 4.89 & 20.5 & 4.82 & 20.8 \\
                         & c & 5.86 & 17.1 & 5.09 & 19.7 & 4.99 & 20.0 \\
\\
\multirow{6}{*}{WO$_3$}  & $\varepsilon$-m & 5.58 & 17.9 & 4.74 & 21.1 & 4.65 & 21.5 \\
                         & tr & 5.54 & 18.1 & 4.68 & 21.4 & 4.56 & 21.9 \\
                         & $\gamma$-m & 5.55 & 18.0 & 4.68 & 21.4 & 4.57 & 21.9 \\
                         & or  & 5.43 & 18.4 & 4.55 & 22.0 & 4.45 & 22.5 \\
                         & t & 6.47 & 15.4 & 5.33 & 18.7 & 5.17 & 19.4 \\
                         & c & 10.09 & 9.90 & 8.27 & 12.1 & 7.98 & 12.5 \\
\end{tabular}
\end{ruledtabular}
\end{table}

Since typically tiny energy differences are involved in transformations between phases with similar stabilities, properly reproducing the stability order is an extremely challenging task, pushing to the limits of accuracy of DFT methods. 
Different functionals may predict different stability sequences, as a consequence of changes in relative total energies of the order of some meV per atom, going from one level of theory to another. 

In the following, we test the performance of the PBE, PBE0 and PBE0$\alpha_{\text{PBE}}$ functionals in this respect. Equilibrium geometries were also investigated, as it is expected that a good xc functional is able to yield optimized lattice parameters within few percent from experiment. Moreover, accurate geometry optimizations are needed for obtaining meaningful total energies to be compared for analysis of phase stability. For this reason, in the present Section all the reported quantities (including dielectric constants, which in turn affect the amount of exact exchange entering the definition of the modified PBE0 functional) were computed at the optimized geometry of the relevant xc functional. Since the \textsc{crystal09} code is very performant when it comes to structural optimizations with hybrid functionals, we used it throughout all the calculations presented in the remaining part of the work.

In Table~\ref{tab4} the electronic dielectric constant and the corresponding exchange fraction calculated at different levels of theory are reported for the 
phases under 
investigation. A partial self-consistency on $\alpha=1/\epsilon_{\infty}$ was carried out by evaluating $\epsilon_{\infty}$ within PBE0$\alpha_{\text{PBE}}$, thus defining the PBE0$\alpha_{\text{PBE}}^{\text{(1)}}$ functional. Comparison of Tables~\ref{tab3} and \ref{tab4} shows the effect of geometry optimization on the computed dielectric constants for the commonly studied polymorphs at the PBE level.

\begin{table}[tb]
\caption{\label{tab5} Differences in total energy per formula unit (in meV, estimated error 1~meV) with respect to a chosen phase (redefining the zero of the energy for each functional individually), for various polymorphs of the studied oxides, computed at different levels of theory. }
\begin{ruledtabular}
\begin{tabular}{lldddd}
& Type & \multicolumn{1}{c}{PBE} & \multicolumn{1}{c}{PBE0} & \multicolumn{1}{c}{PBE0$\alpha_{\text{PBE}}$} & \multicolumn{1}{c}{PBE0$\alpha_{\text{PBE}}^{\text{(1)}}$} \\
\hline
\multirow{2}{*}{ZnO} & wz & 0  & 0  & 0   & 0 \\
                     & zb & 49 & 54 & 102 & 80 \\
\\
\multirow{3}{*}{TiO$_2$} & r & 20 & 62 & 18 & 22 \\
                         & a & 0 & 0 & 0 & 0 \\
                         & b & -5 & 32 & 1 & 2 \\
\\
\multirow{3}{*}{ZrO$_2$} & m & -20 & 17 & 33 & 25 \\
                         & t & 0 & 0 & 0 & 0 \\
                         & c & 74 & 45 & 40 & 36 \\
\\
\multirow{6}{*}{WO$_3$}  & $\varepsilon$-m & 5 & 1 & -2 & -8 \\
                         & tr & 0 & 0 & 0 & 0 \\
                         & $\gamma$-m & 3 & 6 & 2 & 6 \\
                         & or & 6 & 10 & 36 & 1 \\
                         & t & 30 & 70 & -153 & -139 \\
                         & c & 201 & 301 & -403 & -462 \\
\end{tabular}
\end{ruledtabular}
\end{table}

\subsubsection{Polymorphs structures and relative stability}
 
 Apart from wurtzite (wz), which is the most thermodynamically stable phase of ZnO at ambient conditions, the zinc-blende (zb) modification can be stabilized upon growth in specific conditions.\cite{ozgur2005} The rocksalt structure exists only at high pressures, and it is not addressed here. From Table~\ref{tab5} it is seen that the better stability of the wurtzite phase is correctly predicted at all levels of theory, with the partially self-consistent PBE0 confirming the result of the PBE0$\alpha_{\text{PBE}}$. These findings are in agreement with previous calculations using LDA, GGA and screened-exchange hybrid functionals.\cite{uddin2006}

\begin{figure*}[tb]
 \includegraphics[width=0.9\linewidth]{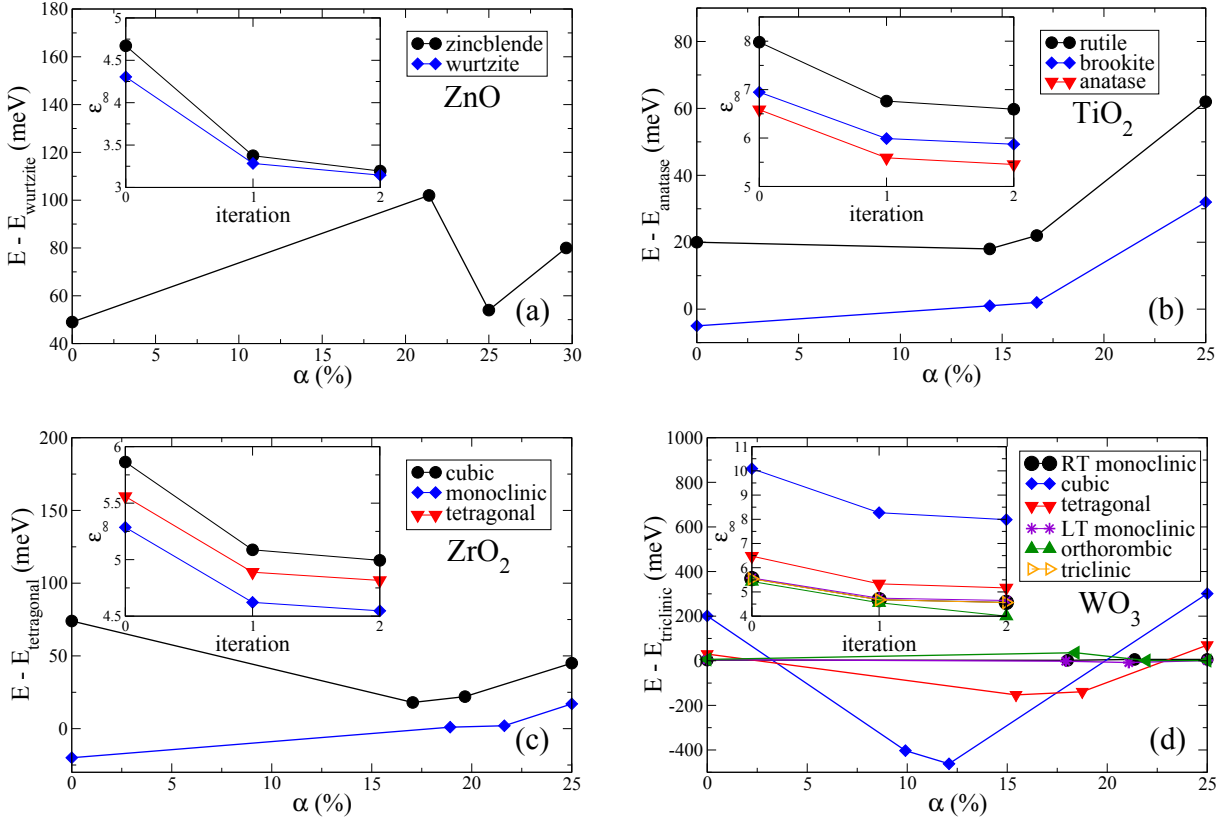}
 \caption{\label{fig1} Variation of total energy as a function of the exchange fraction $\alpha$ entering the definition of the dielectric-dependent hybrid functional. Total energies per unit MO$_x$ (M = Zn, Ti, Zr, W)  are given with reference to that of a chosen polymorph for each material: (a) wurtzite for ZnO, (b) anatase for TiO$_2$, (c) tetragonal for ZrO$_2$, (d) $\gamma$-monoclinic for WO$_3$. In the insets convergence of the electronic dielectric constant $\epsilon_{\infty}$ is analyzed.}
\end{figure*}

The three naturally-occurring polymorphs of TiO$_2$ are rutile (r), anatase (a) and brookite (b). Their relative thermodynamic stability critically depends on crystal size. At ambient conditions, the thermodynamically most stable macrocrystalline (crystal size exceeding 35~nm) phase of TiO$_2$ is rutile. From calorimetric measurements of the phase transformation of anatase and brookite to rutile it is suggested that the phase stability sequence is rutile $>$ brookite $>$ anatase.\cite{kandiel2013} It is well-known that GGA is not able to reproduce the correct sequence, predicting anatase to be more stable than rutile.\cite{muscat2002} Our results confirm such tendency and suggest that, independently of the amount of Hartree-Fock exchange used, hybrid functionals are similarly not able to capture the correct sequence, in agreement with the previous investigation of Labat~\emph{et al.}\cite{labat2007} Recent studies\cite{conesa2010,moellmann2012} demonstrated that dispersion van der Waals 
interactions have to 
be taken 
into 
account in order to correctly reproduce the experimental sequence.

Zirconia features three different structural modifications at ambient pressure; the baddeleyite structure, with a monoclinic (m) unit cell, is stable at room temperature, and transforms into a tetragonal (t) phase at 1480~K, which is stable until 2650~K when it is converted into a cubic (c) flourite phase. In fact, the tetragonal and monoclinic phases can be viewed as distorted cubic structures.\cite{gallino2011} From Table~\ref{tab5} it is inferred that while PBE predicts the correct stability sequence as a function of increasing temperature, PBE0 inverts the monoclinic and tetragonal phases, in fact predicting the latter to be more stable than the former, in agreement with previous studies\cite{gallino2011} but opposite to experimental evidence. Varying the exchange fraction within the dielectric-dependent hybrid scheme did not yield qualitatively different results, similarly to what was found for TiO$_2$.

\begin{table*}[tb]
\caption{\label{tab6} Band gap energy (in eV, estimated error 20~meV) computed within PBE and dielectric-dependent PBE0 for various polymorphs of the studied oxides. Fundamental and direct gaps are reported outside and inside parenthesis, respectively. Calculations were performed within the LCAO scheme at the optimized geometry.}
\begin{ruledtabular}
\begin{tabular}{llccccd}
& Type & PBE & PBE0 & PBE0$\alpha_{\text{PBE}}$ & PBE0$\alpha_{\text{PBE}}^{\text{(1)}}$ 
& \multicolumn{1}{c}{Expt.\footnote{Measured photoemission gap of r-TiO$_2$ from Ref.~\onlinecite{tezuka1994}; direct optical gap of m-, t-, c-ZrO$_2$ from Ref.~\onlinecite{french1994}; optical gap of t- and \mbox{$\gamma$-m-WO$_3$} from Ref.~\onlinecite{ping2014}. For the remaining references on measured band gaps see Table~\ref{tab1}.}} \\
\hline
MgO & rs & 4.63 & 7.38 & 8.24 & 8.67 & 7.83 \\
\\
\multirow{2}{*}{ZnO} & wz & 1.07 & 3.50 & 3.32 & 4.07 & 3.44 \\
                     & zb & 0.91 & 3.29 & 2.93 & 3.76 \\
\\
\multirow{3}{*}{TiO$_2$} & r & 1.71 & 3.90 & 2.76 & 2.96 & 3.3 \\
                         & a & 2.14 (2.23) & 4.36 (4.36) & 3.42 (3.46) & 3.66 (3.70) & 3.42 \\
                         & b & 2.36 & 4.46 & 3.53 & 3.71 & \\
\\
\multirow{3}{*}{ZrO$_2$} & m & 3.38 (3.78) & 5.67 (6.10) & 5.09 (5.59) & 5.34 (5.77) & 5.83 \\
                         & t & 3.80 (3.86) & 6.07 (6.20) & 5.43 (5.53) & 5.66 (5.76) & 5.78 \\
                         & c & 3.12 (3.63) & 5.43 (6.04) & 4.67 (5.25) & 4.91 (5.50) & 6.1 \\
\\
\multirow{6}{*}{WO$_3$}  & $\varepsilon$-m & 1.80 & 3.85 & 3.28 & 3.52 \\
                         & tr & 1.67 & 3.68 & 3.06 & 3.36 \\
                         & $\gamma$-m & 1.62 & 3.67 & 3.07 & 3.34 & 2.6 \\
                         & or & 1.39 & 3.34 & 2.81 & 3.09 \\
                         & t & 0.53 & 2.23 & 1.53 & 1.76 & 1.8 \\
                         & c & 0.55 (1.58) & 2.19 (3.21) & 1.15 (2.17) & 1.29 (2.31) \\
\end{tabular}
\end{ruledtabular}
\end{table*}

Finally, the case of WO$_3$ deserves particular attention, since its phase diagram at ambient pressure is exceptionally rich, exhibiting five different polymorphs transforming into each other by variation of the temperature. The simple cubic (c) structure, only recently successfully stabilized at ambient conditions,\cite{crichton2003} is constituted by a lattice of regular corner-sharing WO$_6$ octahedra, forming a strongly ionic compound. Several structures are successively obtained by decreasing the temperature from the melting point at 1700~K: tetragonal (t, stable above $740\,^{\circ}\text{C}$), orthorombic (or, from $330\,^{\circ}\text{C}$ to $740\,^{\circ}\text{C}$), room-temperature $\gamma$-monoclinic ($\gamma$-m, from $17\,^{\circ}\text{C}$ to $330\,^{\circ}\text{C}$), triclinic (tr, from $-50\,^{\circ}\text{C}$ to $17\,^{\circ}\text{C}$), and low-temperature $\varepsilon$-monoclinic ($\varepsilon$-m, from $-140\,^{\circ}\text{C}$ to $-50\,^{\circ}\text{C}$).\cite{wang2011} 

Octahedra become more and more distorted going 
from the tetragonal to the low-temperature 
monoclinic phase. 
Previous works\cite{dewijs1999,wang2011,ping2014} investigated how such distortions affect the electronic band structure, as a consequence of energy-lowering rearrangement of the WO$_6$ units. The recent study by Ping and Galli\cite{ping2014} suggested that tilting of the WO$_6$ octahedra along different crystalline axis is the main responsible for band gap opening going from the cubic and tetragonal phases (where no tilting is present) to the other, less symmetric structures. This is apparent from Table~\ref{tab6}, in which the reported band gap increases by as much as $\sim 1$~eV or more passing from the tetragonal and cubic structures to the more distorted ones. A similar relationship between structural and electronic properties was found in a recent DFT study on doped WO$_3$,\cite{tosoni2014} in which intercaled atoms drove the structural deformation responsible for the large band gap reduction.

As expected, the electronic dielectric constant follows the opposite trend with respect to the band gap, being larger in the cubic and tetragonal phases and decreasing with increasing band gap, as shown in Table~\ref{tab4}. This in turn affects the corresponding exchange fractions, whose strong dependence upon the phase considered is at the origin of the wrong prediction of the stability order of WO$_3$ within the dielectric-dependent hybrid scheme. In fact, Table~\ref{tab5} shows that, while using standard, fixed-$\alpha$ PBE0 functional the cubic and tetragonal phases were predicted to be the least stable, in agreement with the experimental order as a function of temperature, they turn to be the most stable ones when $\alpha$ is obtained from the computed dielectric constant. This behavior is not observed in the other materials, where the electronic and dielectric properties vary more smoothly upon the phase transition, or, equivalently, the electronic structure correlates more weakly to lattice 
distortions. In those cases, the dielectric-dependent hybrid approach proved to qualitatively reproduce the results of standard PBE0. This is clearly seen in Fig.~\ref{fig1}, where relative total energy is plotted as a function of the exchange fraction. Non-crossing lines on the positive energy half-plane indicate that no phase stability inversion occurs going from 
PBE0 to 
PBE0$\alpha_{\text{PBE}}$, 
contrary to what happens in the case of WO$_3$.

\subsubsection{Band gap dependence on polymorphic structure}

The interesting case of WO$_3$ shows that the electronic structure can be significantly affected by the polymorphic structure. Having assessed the performance of the dielectric-dependent hybrid method in band gaps computation of some well-characterized oxide polymorphs, it is worthwhile to investigate the electronic structure of other less common phases.

Table~\ref{tab6} collects computed band gaps for all the phases studied so far. Here, all calculations were performed at the optimized geometries. Due to the experimental difficulty in obtaining samples of a well-defined phase, measured gaps are not available for all the structures or are very disperse, as in the case of brookite TiO$_2$.\cite{dipaola2013} Hence, comparison with previous hybrid and $\mathit{GW}$ results provides a good benchmark for our calculations. 

In general, the PBE0$\alpha_{\text{PBE}}$ confirms itself as an excellent method for computing accurate electronic structures. Notice however that self-consistency on the dielectric constant does not always improve the computed gap, as happens in the case of MgO and ZnO. Analysis of Table~\ref{tab4} and of the insets in Fig.~\ref{fig1} shows that convergence on the dielectric constant is rapidly achieved, as also reported in Ref.~\onlinecite{skone2014}. Due to the considerable cost of geometry optimizations for the largest 
cells, we limited ourself to a single self-consistency step on $\alpha$, being confident that PBE0$\alpha_{\text{PBE}}^{\text{(1)}}$ results are not far from that of the fully self-consistent PBE0. 

In the following, we briefly compare our results with the existing theoretical and, when available, experimental literature. 

For ZnO, previous hybrid calculations\cite{uddin2006} predicted the band gap in the zinc-blende phase to be $\sim 0.2$~eV smaller than in wurtzite, which is also supported by experimental evidence,\cite{lee2002} confirming our PBE0$\alpha_{\text{PBE}}$ results. 

For TiO$_2$ we obtained a direct gap for rutile which is smaller by $\sim 0.7$~eV than in the anatase phase. Our computed gap for rutile is smaller than previous $\mathit{G_0W_0}$@LDA/GGA results spanning the range from $3.3$ to $3.6$~eV,\cite{kang2010,chiodo2010,landmann2012,zhu2014} but close to the value of $2.85$~eV obtained within $\mathit{G_0W_0}$@LDA+U,\cite{patrick2012} and compatible with the measured QP gap of $3.3\pm 0.5$~eV.\cite{tezuka1994} For the brookite phase, a recent $\mathit{G_0W_0}$@PBE investigation\cite{zhu2014} yielded a direct band gap of $3.86$~eV, larger than in rutile and anatase, and in good agreement with our PBE0$\alpha_{\text{PBE}}^{\text{(1)}}$ calculations. 

For ZrO$_2$, our predicted band gap ordering for the three phases is in line with previous investigations using hybrid functionals\cite{gallino2011} and $\mathit{GW}$,\cite{jiang2010,gruening2010} and numerical values agree very well with the $\mathit{GW_0}$@LDA results of Ref.~\onlinecite{jiang2010} reporting an indirect band gap of $5.34$~eV, $5.92$~eV and $4.97$~eV for the monoclinic, tetragonal and cubic structures, respectively. 

Only recently $\mathit{GW}$ calculations were performed for all the ambient pressure polymorphs of WO$_3$.\cite{ping2014} Computed band gaps within $\mathit{G_0W_0}$@LDA are in good agreement with our PBE0$\alpha_{\text{PBE}}^{\text{(1)}}$ results, and confirm our obtained band gap ordering.

\begin{table}[tb]
\caption{\label{tab7} Optimized cell parameters for various polymorphs of the studied oxide semiconductors, computed at different levels of theory. Calculations were performed within the LCAO scheme using \textsc{crystal09}.}
\begin{ruledtabular}
\begin{tabular}{llccccc} 
& Type & Parameter & PBE & PBE0 & PBE0$\alpha_{\text{PBE}}^{\text{(1)}}$
& Expt.\footnote{See the Supplemental Material.} \\
\hline

MgO & rs & a (\AA{}) & 4.258 & 4.207 & 4.185 & 4.212 \\
\\
\multirow{2}{*}{ZnO} & \multirow{2}{*}{wz} & a (\AA{}) & 3.292 & 3.264 & 3.257 & 3.250 \\
    &     & c (\AA{}) & 5.119 & 5.077 & 5.071 & 5.207 \\
\\
\multirow{2}{*}{TiO$_2$}
& \multirow{2}{*}{a} & a (\AA{}) & 3.821 & 3.765 & 3.789 & 3.781 \\
&                    & c (\AA{}) & 9.672 & 9.656 & 9.631 & 9.515 \\
\\
\multirow{2}{*}{ZrO$_2$}
& \multirow{2}{*}{t} & a (\AA{}) & 3.640 & 3.608 & 3.613 & 3.571 \\
&                    & c (\AA{}) & 5.288 & 5.209 & 5.224 & 5.182 \\
\\
\multirow{4}{*}{WO$_3$}
& \multirow{4}{*}{$\gamma$-m} & a (\AA{}) & 7.444 & 7.334 & 7.348 & 7.306 \\
 &                            & b (\AA{}) & 7.672 & 7.605 & 7.616 & 7.540 \\
 &                            & c (\AA{}) & 7.885 & 7.801 & 7.817 & 7.692 \\
 &                            & $\beta$ ($^{\circ}$) & 90.65 & 90.61 & 90.62 & 90.88 \\
\end{tabular}
\end{ruledtabular}
\end{table}

\subsubsection{Equilibrium geometries}

Finally, in Table~\ref{tab7} we report the optimized cell parameters computed within PBE, standard PBE0 and dielectric-dependent PBE0 for some selected polymorphs. Full optimization of both cell parameters and atomic positions was carried out. Results indicate that dielectric-dependent PBE0 generally performs as well as standard PBE0 when it comes to structural properties, the error with respect to experimental lattice constants being in most cases within $\sim 2\%$.

\begin{table}[b]
\caption{\label{tab8} Relative total energies $E_{\text{tot}}$ (in meV, estimated error 1~meV) per unit Ti$_2$O$_3$, band gap $E_g$ (in eV, estimated error 20~meV) and lattice constants $a$ and $c$ (in \AA{}), computed at the PBE0$\alpha_{\text{PBE0}}^{\text{(1)}}$ level. AFM1 and AFM2 labels two different antiferromagnetic (AFM) configurations; FM refers to the ferromagnetic one. $\alpha=12.8 \%$ was used in the dielectric-dependent PBE0; this value was obtained from the electronic dielectric constant $\epsilon_{\infty} = 7.81 $ computed within PBE0$\alpha_{\text{PBE0}}$ in the AFM1 configuration. Total energies are reported relative to that of the lowest-energy AFM1 phase. For spin configurations the notation (Ti$_1$, Ti$_2$, Ti$_3$, Ti$_4$), with Ti$_n = +/-$, indicates spin up/down for the $3d$ electron of the Ti$_n$ atom in the unit cell (see Fig.~\ref{fig2}). Calculations were performed within the LCAO scheme with \textsc{crystal09} and for the optimized structure.}
\begin{ruledtabular}
\begin{tabular}{lcccccc}
Magnetic configuration & $E_{\text{tot}}$ & $E_g$ & $a$ & $c$ \\
\hline
AFM1 ($+$, $-$, $-$, $+$)  & 0 &  0.59 & 5.167 & 13.671 \\
AFM2 ($+$, $-$, $+$, $-$)  & 6 &  \\
Nonmagnetic                & 13 & \\
FM                         & 13 & \\
Expt.\footnote{Experimental band gap from combined conductivity and thermoelectric coefficient measurements (Ref.\onlinecite{shin1973}). Experimental lattice parameters from Ref.~\onlinecite{rice1977}.}
&    & 0.11 & 5.157 & 13.610 \\ 
\end{tabular}
\end{ruledtabular}
\end{table}

\begin{table*}[tb]
\caption{\label{tab9} Reaction energies (in eV, estimated error 10~meV) for the TiO$_2$ reduction, computed with different hybrid functionals, and compared with experimental heats of reaction. The exchange fraction $\alpha=1/\epsilon_{\infty}$ (\%) used in calculations on TiO$_2$ and Ti$_2$O$_3$ is reported. The computed dielectric constants $\epsilon_{\infty}$ within PBE0 and PBE0$\alpha_{\text{PBE0}}$, respectively, are: 6.10 and 6.65 for TiO$_2$, and 5.63 and 7.81 for Ti$_2$O$_3$ (estimated errors within 0.02). Calculations were performed within the LCAO scheme using \textsc{crystal09}.}
\begin{ruledtabular}
\begin{tabular}{lrcccc} 
\multirow{3}{*}{Reaction} & & PBE0 & PBE0$\alpha_{\text{PBE0}}$ & PBE0$\alpha_{\text{PBE0}}^{\text{(1)}}$ 
& \multirow{3}{*}{Expt.\footnote{Ref.~\onlinecite{lutfalla2011} and references therein.}} \\
 & $\alpha$(TiO$_2$) & 25  & 16.4 & 15.0 \\
 & $\alpha$(Ti$_2$O$_3$) & 25 & 17.8 & 12.8 \\
\hline
 $2\,\text{TiO}_2 + \text{H}_2 \rightarrow \text{Ti}_2\text{O}_3 + \text{H}_2\text{O}$ & & 0.96 & 1.22 & 1.26 & 1.30 \\
 $2\,\text{TiO}_2 \rightarrow \text{Ti}_2\text{O}_3 + \frac{1}{2} \text{O}_2$
 & & 3.55 & 3.81 & 3.85 & 3.81 \\
\end{tabular}
\end{ruledtabular}
\end{table*}

\subsection{Stoichiometry transformations: the chemical reduction of TiO$_2$}

We finally assess the performance of the dielectric-dependent hybrid method in calculating reaction energies for the chemical reduction of TiO$_2$ to Ti$_2$O$_3$. Differences in total energy involved in chemical reactions are typically of the order of few eV, compared to few meV or tens of meV, observed in structural phase transitions of the previous section. This allows us to meaningfully address the issue of whether the method is able or not to \emph{quantitatively} improve over standard hybrid functionals in terms of computed total energies.

We are concerned here with the two reduction reactions of TiO$_2$:

\begin{gather}
 2\,\text{TiO}_2 + \text{H}_2 \rightarrow \text{Ti}_2\text{O}_3 + \text{H}_2\text{O} 
 \label{eq:reduction1} \\
 2\,\text{TiO}_2 \rightarrow \text{Ti}_2\text{O}_3 + \frac{1}{2} \text{O}_2.
 \label{eq:reduction2}
\end{gather}

\begin{figure}[b]
 \includegraphics[width=0.6\linewidth]{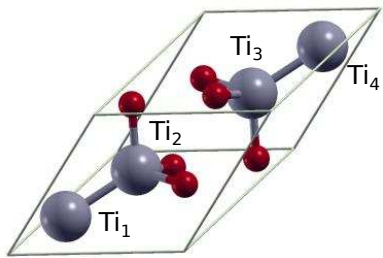}
 \caption{\label{fig2} Bulk primitive cell of corundum Ti$_2$O$_3$. Grey and red spheres represent Ti and O atoms, respectively.}
\end{figure}

Ti$_2$O$_3$ is a small gap semiconductor with a corundum structure. Its bulk unit cell is rhombohedral and contains two Ti$_2$O$_3$ units, with atoms positioned as shown in Fig.~\ref{fig2}. The magnetic properties of Ti$_2$O$_3$ at low temperature have been subject of an intensive investigation both at the experimental\cite{abrahams1963,kendrick1968} and theoretical level.\cite{catti1997,hu2011,iori2012,guo2012} However, no general consensus has been achieved on whether the low-temperature ground state of Ti$_2$O$_3$ is antiferromagnetic (AFM) or nonmagnetic. We re-investigated the issue using both standard and dielectric-dependent PBE0. Standard non-hybrid DFT predicts the material to be metallic,\cite{mattheiss1996} at odds with the experimental evidence. Our spin-polarized PBE0 calculations yielded an AFM ground state with AFM1 ($+$, $-$, $-$, $+$) spin configuration of Ti $3d$ electrons. Previous studies within Hartree-Fock,\cite{catti1997} hybrid DFT and DFT+U\cite{hu2011} reported the same result. 
Ferromagnetic (FM) and nonmagnetic solutions are found to be much higher in energy at the PBE0 level. Interestingly, different investigations using screened-exchange hybrid functionals predicted the nonmagnetic configuration to be the most stable.\cite{iori2012,guo2012}
Calculations using the dielectric-dependent PBE0$\alpha_{\text{PBE0}}$ functional confirmed the magnetic configuration to be the AFM1, as reported in Table~\ref{tab8}. The corresponding computed band gap of $0.59$~eV is larger than the experimental value of $0.11$~eV obtained by combined conductivity and thermoelectric coefficient measurements,\cite{shin1973} but still closer than our PBE0 result of $2.63$~eV, and in agreement with screened exchange hybrid calculations,\cite{iori2012} reporting a value of $0.57$~eV. 

For TiO$_2$, we considered the rutile phase in the present analysis, since experimental data are available for the reaction enthalpy of the transformation to Ti$_2$O$_3$. We carried out full geometry optimizations within PBE0, PBE0$\alpha_{\text{PBE0}}$, and partially self-consistent PBE0$\alpha_{\text{PBE0}}^{\text{(1)}}$. A band gap of $3.03$~eV was obtained within PBE0$\alpha_{\text{PBE0}}^{\text{(1)}}$, which is very close to our previously reported PBE0$\alpha_{\text{PBE}}^{\text{(1)}}$ value of $2.96$~eV (see Table~\ref{tab6}), confirming that the self-consistency is practically achieved.

The computed energies of reactions \eqref{eq:reduction1} and \eqref{eq:reduction2} are summarized in Table~\ref{tab9}.\footnote{Ground-state total energy calculations on molecules were carried out at the PBE0 level, using optimized basis sets for molecular calculations as described in Section~\ref{sec:comp-det}, aiming at accurately reproducing experimental atomization energies. In fact, using an oxygen basis set optimized for crystal calculations resulted in poor agreement with experiment in terms of atomization energies of O$_2$ and H$_2$O.} The dielectric-dependent PBE0 yielded results in excellent agreement with experiment, with a clear improvement over standard PBE0. Previous first-principles investigations within GGA+U\cite{hu2011,lutfalla2011} found that choosing the value of the Hubbard parameter $U$ between 2 and 3~eV results in reaction energies close to experiment. However, the chosen value of $U$ strongly affects the computed electronic band structure. Ti$_2$O$_3$ was reported to be metallic at 
$U=2$~eV, while $U=3$~eV was needed to open a gap of comparable size to the experimental one. For TiO$_2$ it was demonstrated\cite{persson2005} that $U$ should be set to $\sim 10$~eV in order to obtain a band gap of 3~eV. Another deficiency of the DFT+U approach lies in the failure to predict accurate lattice parameters for Ti$_2$O$_3$.\cite{hu2011} Both of the above issues are elegantly solved within the dielectric-dependent hybrid method, which we found out to be able to accurately characterize reaction energies, electronic structures and structural parameters within a single, parameter-free, and fully \emph{ab initio} theoretical approach.

\section{Conclusions}

We have reported a systematic analysis of band gaps, equilibrium geometries and phase stability of several ambient-pressure bulk polymorphs of wide gap oxide semiconductors of current interest for energy applications. Hybrid xc functionals with a fraction of Hartree-Fock exchange weighted on the static electronic dielectric constant have been confirmed to provide an improvement over standard hybrid functionals concerning band gap estimation, the resulting accuracy being comparable to the more expensive $\mathit{GW}$ method. Equilibrium geometries computed within the dielectric-dependent hybrid scheme have been found to be as accurate as with standard hybrids. Predicted phase stability sequences have also been found to be the same with the two methods, with the exception of WO$_3$, in which a peculiar correlation between structure and band gap leads to different predicted stabilities. The stoichiometry transformation of TiO$_2$ to Ti$_2$O$_3$ has then been investigated. The dielectric-dependent hybrid method 
has proved successful in describing band gaps and equilibrium geometries of both materials, while providing reaction energies for the TiO$_2$ reduction in excellent agreement with experiments. In conclusion, the method qualifies itself for accurate determination of either excited- and ground-state properties of bulk solids, constituting a useful tool in those situations in which an accurate description of both is required within a common theoretical framework.

\begin{acknowledgments}

The $\mathit{GW}$ calculations have been performed at the CINECA
HPC facility through the  IsC12\_CDSBiT, PoliM\_EneDAF and  UniMI\_FisMaF allocations.
CDV and GP thank the Italian MIUR for the FIRB Project RBAP115AYN, the Cariplo Foundation for the grant n. 2013-0615 and the the COST Action CM1104.
GO and LC acknowledge the ETSF-Italy\cite{etsf} for computational support.
MG acknowledges useful interactions with Jonathan H. Skone and Fenggong Wang.

\end{acknowledgments}

\end{document}